\documentclass[aps,pre,10pt,twocolumn,a4paper,superscriptaddress,nofootinbib]{revtex4-1}
\usepackage[utf8]{inputenc}
\usepackage{amsmath}
\usepackage{amssymb}
\usepackage{graphicx}
\usepackage{color}

\DeclareMathOperator{\arctanh}{arctanh}

\renewcommand{\Re}{\mathop\mathrm{Re}\nolimits}
\renewcommand{\Im}{\mathop\mathrm{Im}\nolimits}

\begin{document}

\title{Surface density of states in superconductors with inhomogeneous pairing constant: Analytical results}

\author{Ya.~V.\ Fominov}
\affiliation{L.~D.\ Landau Institute for Theoretical Physics RAS, 142432 Chernogolovka, Russia}
\affiliation{Moscow Institute of Physics and Technology, 141700 Dolgoprudny, Russia}
\affiliation{National Research University Higher School of Economics, 101000 Moscow, Russia}

\author{A.~A.\ Mazanik}
\affiliation{Moscow Institute of Physics and Technology, 141700 Dolgoprudny, Russia}
\affiliation{BLTP, JINR, 141980 Dubna, Russia}

\author{M.~V.\ Razumovskiy}
\affiliation{Moscow Institute of Physics and Technology, 141700 Dolgoprudny, Russia}

\begin{abstract}
We consider a superconductor with surface suppression of the BCS pairing constant $\lambda(x)$.
We analytically find the gap in the surface density of states (DOS), behavior of the DOS  $\nu(E)$ above the gap, a ``vertical'' peculiarity of the DOS around an energy equal to the bulk order parameter $\Delta_0$, and a perturbative correction to the DOS at higher energies.
The surface gap in the DOS is parametrically different from the surface value of the order parameter due to a difference between the spatial scale $r_c$, at which $\lambda(x)$ is suppressed, and the coherence length. The vertical peculiarity implies an infinite-derivative inflection point of the DOS curve at $E=\Delta_0$ with square-root behavior as $E$ deviates from $\Delta_0$. The coefficients of this dependence are different at $E<\Delta_0$ and $E>\Delta_0$, so the peculiarity is asymmetric.
\end{abstract}

\date{26 December 2019}

\maketitle

\tableofcontents

\section{Introduction}
\label{sec:intro}

The standard BCS theory of superconductivity assumes a homogeneous pairing constant $\lambda(\mathbf r) = \lambda_0$ that leads to the formation of Cooper pairs and the superconducting condensate characterized by the order parameter  $\Delta(\mathbf r) = \Delta_0$ \cite{Bardeen1957b,TinkhamBook}.
Obviously, the inhomogeneous pairing constant should essentially influence the superconducting properties of the system.
One of the basic examples of inhomogeneous $\lambda(\mathbf r)$ dependence is a hybrid superconductor/normal-metal (SN) structure, in which $\lambda=0$ in the N part. The corresponding inhomogeneity of $\Delta(\mathbf r)$ gives rise to a prominent effect of Andreev reflection at the SN interface and to possibility of Andreev bound states in the N section of the structure \cite{[][{ [Sov. Phys. JETP \textbf{19}, 1228 (1964)].}]Andreev1964RusEng,TinkhamBook}.

In the absence of interfaces, effects of local inhomogeneities of the pairing constant on the quasiparticle density of states (DOS) and other physical quantities have been studied in bulk (clean) superconductors with both conventional $s$-wave and anisotropic $d$-wave pairing \cite{Shnirman1999.PhysRevB.60.7517,Andersen2006.PhysRevLett.96.097004,Bespalov2019.PhysRevB.100.094507}. Although details depend on the specific type of system under discussion, the modifications of the DOS are generally related to the formation of the Andreev bound states at inhomogeneities of $\lambda(\mathbf r)$ and $\Delta(\mathbf r)$. Periodic-in-space modulations of $\lambda(\mathbf r)$ influence basic superconducting properties such as the critical temperature and the energy gap \cite{Martin2005.PhysRevB.72.060502,Zou2008.PhysRevB.77.144523}. Inhomogeneous pairing also influences superconducting properties in non-BCS models \cite{Romer2012.PhysRevB.86.054507}.

An inhomogeneous spatial profile of $\Delta(\mathbf r)$ represents the Andreev potential well. While in clean superconductors this results in the Andreev bound states, in the diffusive limit, the discrete Andreev levels are effectively smeared out and a spectral gap is formed instead. This spectral gap $E_g$ is a functional of the full $\Delta(\mathbf r)$ profile and marks the minimal energy of a continuum quasiparticle spectrum (see examples of the $E_g$ calculation in Refs.\ \cite{[][{ [Sov. Phys. JETP \textbf{69}, 805 (1989)].}]Golubov1989RusEng,Zhou1998}).

In a conventional $s$-wave superconductor, a surface by itself does not cause pair breaking. Theoretically, if a surface simply defines the geometry of a sample, the order parameter and the
DOS do not vary in space, i.e., the bulk solution is valid everywhere inside the superconductor and is not distorted by the surface \cite{TinkhamBook}.

At the same time, in realistic samples, the surface can be imperfect in the sense that it influences superconductivity due to additional effects such as thin oxide layers, absorbed impurities, deviations from stoichiometry, etc. \cite{AntoineBook,Gurevich2012}. Surface properties can also be manipulated on purpose by chemical treatment or by irradiation \cite{Halama1971}. A theoretical description of those effects is complicated and definitely nonuniversal. To model suppression of superconductivity near the surface, one can assume surface suppression of the BCS pairing constant $\lambda (\mathbf r)$ \cite{MazanikBSThesis2016,RazumovskiyBSThesis2017,Gurevich2017,Kubo2019}. Microscopically, this effect can be due to changes in lattice properties (i.e., phonons) or in electron-phonon interaction in the vicinity of an imperfect surface.
Moreover, even in ideal samples, the near-surface pairing constant can be suppressed due to properties of surface phonons \cite{Noffsinger2010.PhysRevB.81.214519} (note at the same time that the opposite effect of surface enhancement of superconductivity has also been discussed \cite{Ginzburg1964}).

In this paper, we study the surface DOS in a diffusive superconductor with the pairing constant $\lambda (\mathbf r)$ varying near the surface.
The surface DOS can be directly probed by scanning tunneling spectroscopy and also directly influences the surface impedance (in particular, its real part, the surface resistance) \cite{TinkhamBook,Gurevich2017,Kubo2019}.

\begin{figure}
 \center{\includegraphics[width=0.8\columnwidth]{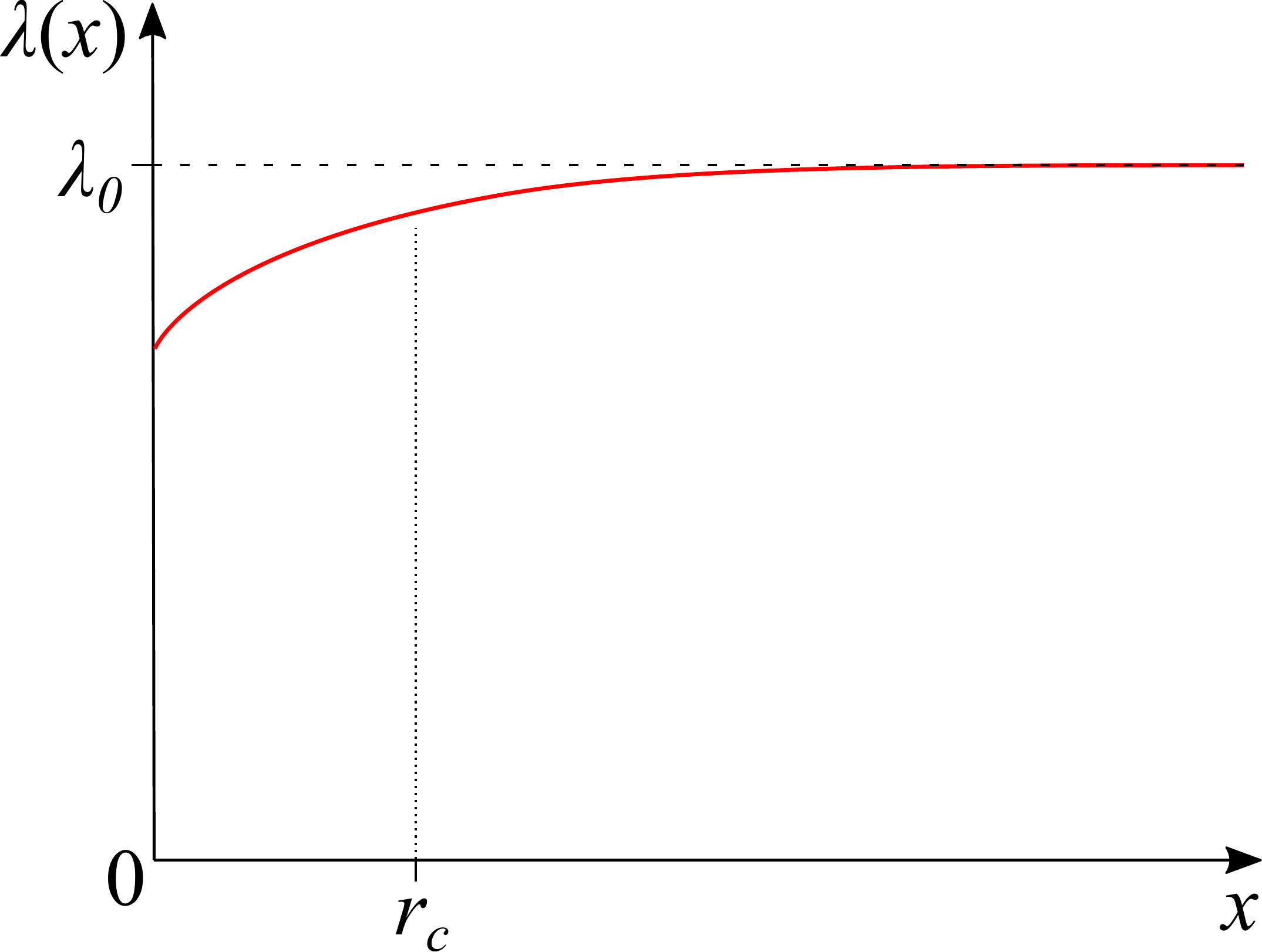}}
\caption{Surface suppression of the BCS pairing constant $\lambda(x)$ (schematic plot). The superconductor occupies the $x>0$ region. The bulk value is denoted $\lambda_0$. The surface suppression takes place near the surface, at $x\sim r_c$. The same model was considered in Refs.\ \cite{MazanikBSThesis2016,RazumovskiyBSThesis2017,Gurevich2017}.}
\label{fig:system}
\end{figure}

A complementary problem of the DOS in superconductors with \emph{random} $\lambda (\mathbf r)$ has been studied before by Larkin and Ovchinnikov \cite{[][{ [Sov. Phys. JETP  \textbf{34}, 1144 (1972)].}]Larkin1971RusEng} and in subsequent publications \cite{Meyer2001,[][{ [JETP  \textbf{117}, 487 (2013)].}]Skvortsov2013RusEng}. In contrast, similarly to Gurevich and Kubo \cite{Gurevich2017}, we assume deterministic form of the  $\lambda (\mathbf r)$ dependence; see Fig.~\ref{fig:system}.
In Ref.\ \cite{Gurevich2017}, the analytical approach to calculating the DOS in the model of Fig.~\ref{fig:system} was formulated and numerical results for the surface DOS were presented. In this paper, we mainly focus on analytical results for the surface DOS. In particular, we analyze the suppression of the gap edge $E_g$ (with respect to the bulk value of the order parameter $\Delta_0$) and behavior of the DOS above $E_g$. We also demonstrate peculiar DOS behavior in the vicinity of $E=\Delta_0$.

The paper is organized as follows:
In Sec.~\ref{sec:general}, we formulate equations of the self-consistent quasiclassical theory in the diffusive limit.
In Sec.~\ref{sec:self-cons_perturb}, we recall how self-consistency for the order parameter is taken into account in the case of small disturbance of $\lambda(\mathbf r)$.
In Sec.~\ref{sec:DOSperturbative}, we analyze the perturbative regime of energies $E>\Delta_0$.
In Sec.~\ref{sec:DOSnonperturb}, we consider the nonperturbative regime of $E\sim \Delta_0$; this section contains our main results for the gap $E_g$ and the behavior of the DOS at $E\approx E_g$ and $E\approx \Delta_0$.
In Sec.~\ref{sec:numerical}, we illustrate and discuss our results.
In Sec.~\ref{sec:conclusions}, we present our conclusions.
Finally, some details of calculations are presented in the Appendixes.

Throughout the paper, we employ the units with $k_B = \hbar = 1$.

\section{General equations}
\label{sec:general}

To calculate the DOS in a diffusive inhomogeneous superconductor, we employ the quasiclassical approach \cite{Usadel1970,Belzig1999}. With the help of the standard $\theta$ parametrization, we can write the normal and anomalous Green functions as $G=\cos\theta$ and $F=\sin\theta$, respectively. The coupled system of the Usadel equation \cite{Usadel1970} and the self-consistency equation can then be written as
\begin{gather}
\frac D2 \nabla^2 \theta(\omega_n, \mathbf r) - \omega_n \sin\theta(\omega_n, \mathbf r) + \Delta(\mathbf r)  \cos\theta(\omega_n, \mathbf r) =0,
\label{Usadel}
\\
\Delta(\mathbf r) = \lambda(\mathbf r) \pi T \sum_n \sin\theta(\omega_n, \mathbf r).
\label{Delta(r)}
\end{gather}
Here $D$ is the diffusion constant, $T$ is temperature, $\omega_n = \pi T (2n+1)$ is the Matsubara frequency, and $\Delta$ is the superconducting order parameter.

We consider a superconductor with a flat surface, so that all quantities depend only on $x$, the coordinate along the normal to the surface (the superconductor occupies the $x>0$ half space).
To complete the system of equations, we must take into account the boundary condition at the surface,
\begin{equation} \label{bc}
\partial \theta / \partial x \bigr|_{x=0} =0.
\end{equation}

Our model of the surface suppression of superconductivity is defined by the form of the $\lambda(x)$ dependence (similarly to Ref.\ \cite{Gurevich2017}). At $x\to\infty$, the pairing constant $\lambda(x)$ tends to its bulk value $\lambda_0$, while we assume it to vary near the surface at some characteristic length scale $r_c$;
see Fig.~\ref{fig:system}.

The DOS at each point (normalized to the normal-metallic value $\nu_0$) can be calculated from the normal Green function after analytical continuation to real energies $E$:
\begin{equation} \label{DOS}
\left. \frac{\nu(E, x)}{\nu_0} =\Re \cos\theta(\omega_n, x) \right|_{\omega_n\mapsto -i(E+i0)}.
\end{equation}
The DOS in our problem is an even function of energy, so below we discuss only $E>0$.

One could reformulate Eqs.\ (\ref{Usadel})--(\ref{bc}) in the real-energy representation from the very beginning. However, we prefer to start from the Matsubara representation since it is convenient for treating the self-consistency equation (\ref{Delta(r)}) (no singularities in the anomalous Green function under the sum) and switch to real $E$ only in the end of calculation, according to Eq.\ (\ref{DOS}).

The solution of the system of Eqs.\ (\ref{Usadel})--(\ref{bc}) is inhomogeneous only because of the $\lambda (x)$ dependence. In the case of $\lambda(x) \equiv \lambda_0$, the bulk solution would be valid everywhere in the superconductor up to the surface. This bulk solution yields
\begin{equation} \label{bulksolution}
\cos\theta_0 (\omega_n) = \frac{\omega_n}{\sqrt{\omega_n^2 + \Delta_0^2}},
\qquad
\sin\theta_0 (\omega_n) = \frac{\Delta_0}{\sqrt{\omega_n^2 + \Delta_0^2}},
\end{equation}
and Eq.\ (\ref{DOS}) then immediately produces the BCS DOS,
\begin{equation} \label{DOS_BCS}
\frac{\nu_\mathrm{BCS}(E)}{\nu_0} =\Re  \frac{E}{\sqrt{E^2 - \Delta_0^2}}.
\end{equation}

\section{Self-consistent perturbation theory}
\label{sec:self-cons_perturb}

Small spatially-dependent inhomogeneities in $\lambda(\mathbf r)$ generate small inhomogeneities in $\Delta(\mathbf r)$ and $\theta(\omega_n, \mathbf r)$:
\begin{align}
\lambda &= \lambda_0 + \lambda_1(\mathbf r),
\\
\Delta &= \Delta_0 + \Delta_1(\mathbf r),
\\
\theta &= \theta_0(\omega) + \theta_1(\omega_n, \mathbf r).
\end{align}
Expanding the Usadel equation (\ref{Usadel}) with respect to small inhomogeneities, we find
\begin{equation} \label{theta1}
\theta_1(\omega_n,\mathbf k) = \Delta_1(\mathbf k) \frac{\cos\theta_0}{\frac D2 k^2 +\omega_n \cos\theta_0 +\Delta_0 \sin\theta_0},
\end{equation}
and the self-consistency equation (\ref{Delta(r)}) yields \cite{[][{ [Sov. Phys. JETP  \textbf{34}, 1144 (1972)].}]Larkin1971RusEng,[][{ [JETP  \textbf{117}, 487 (2013)].}]Skvortsov2013RusEng}
\begin{equation} \label{Delta1}
\frac{\Delta_1(\mathbf k)}{\Delta_0} = L_0(k) \frac{\lambda_1(\mathbf k)}{\lambda_0^2}
\end{equation}
(the combination $|\lambda_1| / \lambda_0^2$ naturally arises as variation of $1/\lambda$).
Here $L_0(k)$ is the static propagator of superconducting fluctuations; see Eq.\ (\ref{L0inv}) in Appendix~\ref{sec:app:L0} for the definition.
This function is real (positive) and even. The behavior of $L_0(k)$ in some limiting cases is considered in Appendix~\ref{sec:app:L0}.

A given form of $\lambda(\mathbf k)$ thus directly determines $\Delta_1(\mathbf k)$ according to the general relation (\ref{Delta1}). Although the characteristic scale for $L_0(k)$ is inverse coherence length, at this scale the decay law only changes to a very slow $1/\ln(k^2)$ form \cite{[][{ [Sov. Phys. JETP  \textbf{34}, 1144 (1972)].}]Larkin1971RusEng}. This decay law cannot lead to convergence of integration when we transform Eq.\ \eqref{Delta1} to coordinate space, so the characteristic scale for $\Delta_1(x)$ is eventually the same as for $\lambda_1(x)$, i.e., it is given by $r_c$ \cite{[][{ [Sov. Phys. JETP  \textbf{34}, 1144 (1972)].}]Larkin1971RusEng}.

It is most convenient to treat relation \eqref{Delta1} within the framework of the Matsubara technique [summation over the Matsubara frequencies is contained in the expression for $L_0(k)$].
The correction to the Green functions [encoded in the correction to the spectral angle $\theta_1(\omega_n,\mathbf k)$] is then immediately given by Eq.\ (\ref{theta1}).
Finally, we need to calculate the DOS according to Eq.\ (\ref{DOS}). This final step must be done at real energies, so there will be a problem at $E\approx \Delta_0$ due to the BCS singularity in the unperturbed Green functions.
The above perturbative approach therefore works only at $E$ above (and not too close to) $\Delta_0$.

\section{Density of states: perturbative regime, \texorpdfstring{$E>\Delta_0$}{E>Delta0}}
\label{sec:DOSperturbative}

The perturbation theory, Eqs.\ (\ref{theta1})--(\ref{Delta1}), immediately produces
\begin{equation}
\left. \frac{\nu_1(E,x)}{\nu_0} = -\Re \left[ \theta_1(\omega_n, x) \sin\theta_0 (\omega_n) \right] \right|_{\omega_n \mapsto -iE}
\end{equation}
for deviation of the DOS from the BCS result, Eq.\ \eqref{DOS_BCS}.

The given function $\lambda_1(x)$ is real and defined at $x>0$. We can symmetrically continue it to the whole axis obtaining an even function.
The Fourier transform can then be written as $\lambda_1(k) = \int dx \cos (kx) \lambda_1(x)$; it is also real and even. We then find the result for the DOS:
\begin{equation} \label{DOSperturb}
\frac{\nu_1(E,x)}{\nu_0} = -\frac{\Delta_0^2}{\lambda_0^2} \frac{E}{\Delta_0^2-E^2} \Im \int\limits_{-\infty}^\infty \frac{dk}{2\pi} \frac{e^{ikx} L_0(k) \lambda_1(k)}{\frac{Dk^2}2 +\sqrt{\Delta_0^2-E^2}}.
\end{equation}
The integral $\int dk e^{ikx} (\dots)$ can be written as $\int dk \cos(kx) (\dots)$, and the result is manifestly zero at $E<\Delta_0$. Of course, the actual local DOS in the inhomogeneous case can be finite at $E<\Delta_0$, however, this region is ``nonperturbative'' from the point of view of our straightforward perturbation theory. This approach only works well at $E>\Delta_0$ (not too close to $\Delta_0$).

The general perturbative result (\ref{DOSperturb}) simplifies considerably if $\lambda_1(k)$ is a decaying function with small characteristic scale so that the integral in Eq.\ (\ref{DOSperturb}) converges at this scale. Physically, this means that $\lambda (x)$ varies slowly enough so that the DOS in this case has the BCS form corresponding to the local value of $\Delta(x)$
\cite{[][{ [Sov. Phys. JETP  \textbf{34}, 1144 (1972)].}]Larkin1971RusEng}. Equation (\ref{DOSperturb}) then yields
\begin{equation} \label{nu1}
\frac{\nu_1(E,x)}{\nu_0} = \frac{\Delta_0^2}{\lambda_0^2} L_0(0) \lambda_1(x) \Re \frac{E}{(E^2-\Delta_0^2)^{3/2}},
\end{equation}
and the same result is obtained directly by varying the BCS expression (\ref{DOS_BCS}) and taking into account Eq.\ (\ref{Delta1}).

At zero temperature, $L_0(0)=1$, and at $E>\Delta_0$ we obtain
\begin{equation} \label{nu1T=0}
\frac{\nu_1(E,x)}{\nu_0} = \frac{E \Delta_0^2}{(E^2-\Delta_0^2)^{3/2}} \frac{\lambda_1(x)}{\lambda_0^2}.
\end{equation}
The same result is obtained directly by varying the BCS expression (\ref{DOS_BCS}) and taking into account the BCS relation $\Delta_0 = 2\omega_D e^{-1/\lambda_0}$ at $T=0$ (here $\omega_D$ is the Debye frequency).

From now on, we consider the case $T=0$, in order to maximize characteristic energy scales related to superconductivity.

The coherence length
\begin{equation}
\xi_0 = \sqrt{\frac D{2\Delta_0}}
\end{equation}
sets the characteristic scale for the fluctuation propagator $L_0(k)$ [at the same time, as we have mentioned above, at $k\gtrsim \xi_0^{-1}$ the $L_0(k)$ function decays very slowly].
At the same time, the denominator in the integral in Eq.\ (\ref{DOSperturb}) varies at $k\sim \xi_E^{-1}$, where the scale is set by a different, energy-dependent coherence length,
\begin{equation} \label{xiE}
\xi_E = \sqrt{\frac D{2 \sqrt{| \Delta_0^2-E^2 |}}}.
\end{equation}

The physical picture beyond the perturbative results (\ref{nu1}) and (\ref{nu1T=0}) is that the DOS adiabatically follows variations of $\Delta(x)$ and has the BCS form corresponding to the local value of the order parameter. This result is valid if $r_c$ exceeds
both $\xi_0$ and $\xi_E$
[slow $\lambda (x)$ function] and reproduces the result for the case of inhomogeneities of large size, obtained by Larkin and Ovchinnikov
\cite{[][{ [Sov. Phys. JETP  \textbf{34}, 1144 (1972)].}]Larkin1971RusEng}.

The calculated DOS at $E>\Delta_0$ is valid at any $x$, in particular, at the surface.
However, the perturbative results (\ref{nu1}) and (\ref{nu1T=0}) become invalid at $E\to \Delta_0$ due to divergence in the denominators (and breakdown of the requirement $r_c > \xi_E$).

At the same time,
we are mainly interested in calculating the surface DOS near $\Delta_0$ and below. In particular, we want to find the shift of the spectrum edge due to inhomogeneity.
This region of energies is nonperturbative and should be treated differently.

\section{Density of states: nonperturbative regime, \texorpdfstring{$E\sim \Delta_0$}{E sim Delta0}}
\label{sec:DOSnonperturb}

Now we assume short-range variation of the pairing constant, so that
\begin{equation} \label{rcllxi0}
r_c \ll \xi_0.
\end{equation}

We substitute $\theta=\pi/2 +i\psi$ (this is convenient for finding the energy gap since $\psi$ is real below the gap).
Introducing dimensionless energy, order parameter (its inhomogeneous part), and coordinate according to
\begin{equation}
\varepsilon = E/\Delta_0,
\qquad
\delta_1(x)= \Delta_1(x)/\Delta_0,
\qquad
X = x/\xi_0,
\end{equation}
we rewrite the Usadel equation \eqref{Usadel} in the real-energy representation as
\begin{equation} \label{UsShG}
\psi'' - \varkappa^2 \sinh( \psi -\psi_0) = \delta_1(X) \sinh \psi,
\end{equation}
where
\begin{equation}
\varkappa = (1-\varepsilon^2)^{1/4}.
\qquad
\psi_0 = \arctanh \varepsilon.
\end{equation}
Here $\psi_0$ is the bulk solution [the real-energy counterpart of Eq.\ \eqref{bulksolution}].
Note that in terms of $\varkappa$, the energy-dependent coherence length \eqref{xiE} can be written as $\xi_E = \xi_0/|\varkappa|$.

At $\varepsilon \sim 1$, we have either $|\varkappa|<1$ or $|\varkappa| \sim 1$.
The characteristic spatial scale for $\psi(X)$, which is determined by $|\varkappa|^{-1}$, is then much larger than
$r_c/\xi_0$, the characteristic spatial scale for $\delta_1(X)$.
The right-hand side (r.h.s.) of Eq.\ \eqref{UsShG} therefore acts as a $\delta$ function and can be taken into account as an effective boundary condition \cite{RazumovskiyBSThesis2017,Gurevich2017}. For that, we integrate Eq.\ (\ref{UsShG}) from 0 to $X_0$, such that $r_c/\xi_0 \ll X_0 \ll |\varkappa|^{-1}$. This scale is small for $\psi(X)$ and large for $\delta_1(X)$. As a result, we obtain the following effective problem:\footnote{Deriving Eq.\ (\ref{d1bound}), we assume
$X_0 \varkappa^2 \sinh(\psi(0)-\psi_0) \ll \psi'(0)$.
As follows from our further calculations, this condition is most restrictive near the gap, where we have $(\psi(0)-\psi_0)\sim 1$ and $\psi'(0) \sim \varkappa$, so our assumption reduces to $X_0 \ll \varkappa^{-1}$. This allows us to choose $X_0$ in the desired range between $r_c/\xi_0$ and $\varkappa^{-1}$.
}
\begin{gather}
\psi'' - \varkappa^2 \sinh( \psi -\psi_0) =0, \label{UsShG0}
\\
\psi'(0) = -d_1 \sinh \psi(0), \label{d1bound}
\end{gather}
where
\begin{equation} \label{d1}
d_1 = -\int_0^\infty \delta_1(X) dX.
\end{equation}
Since $|\delta_1(X)|$ does not exceed unity and the characteristic scale of integration in Eq.\ \eqref{d1} is set by $r_c/\xi_0$, due to condition \eqref{rcllxi0} we have
\begin{equation} \label{d1ll1}
d_1 \ll 1.
\end{equation}

Expressing $d_1$ in terms of $\lambda_1$ with the help of Eq.\ (\ref{Delta1}) and taking into account condition \eqref{rcllxi0}, we find
\begin{equation} \label{d_1}
d_1 = -\frac{L_0(0)}{\lambda_0^2 \xi_0} \int_0^\infty \lambda_1(x) dx.
\end{equation}

Equation (\ref{UsShG0}) is solved by
\begin{equation} \label{psisol}
\psi = 4\arctanh\left( a e^{-\varkappa X} \right) +\psi_0,
\end{equation}
where $a$
should be determined from the boundary condition (\ref{d1bound}):
\begin{equation} \label{d1kappaunsimplified}
\frac{\varkappa^3}{d_1} =
\frac{4a(1+a^2) + \sqrt{1-\varkappa^4} (1+6a^2+a^4)}{4 a (1-a^2)}.
\end{equation}
This equation was derived in Ref.\ \cite{RazumovskiyBSThesis2017} and (in different notations) in Ref.\ \cite{Gurevich2017}.

Finally, the DOS \eqref{DOS} is given by
\begin{equation} \label{DOSpsi}
\frac{\nu(E,x)}{\nu_0} = \Im \sinh\psi(E,x).
\end{equation}

In the following, we analyze the DOS assuming that $\varepsilon$ is close to 1, so that
\begin{equation} \label{kappasmall}
\varkappa \approx \left( 2 (1-\varepsilon) \right)^{1/4},
\qquad
|\varkappa| \ll 1
\end{equation}
(see Appendix~\ref{sec:app:applnonpert}, where the applicability conditions are formulated in terms of the input parameters of our model).
We can then replace the square root in the numerator of Eq.\ \eqref{d1kappaunsimplified} by 1, obtaining the simplified equation
\begin{equation} \label{d1kappa}
\frac{\varkappa^3}{d_1} = \frac{(1+a)^3}{4a(1-a)}.
\end{equation}

\subsection{Energy gap \texorpdfstring{$E_g$}{Eg}}

\begin{figure}
 \center{\includegraphics[width=0.8\columnwidth]{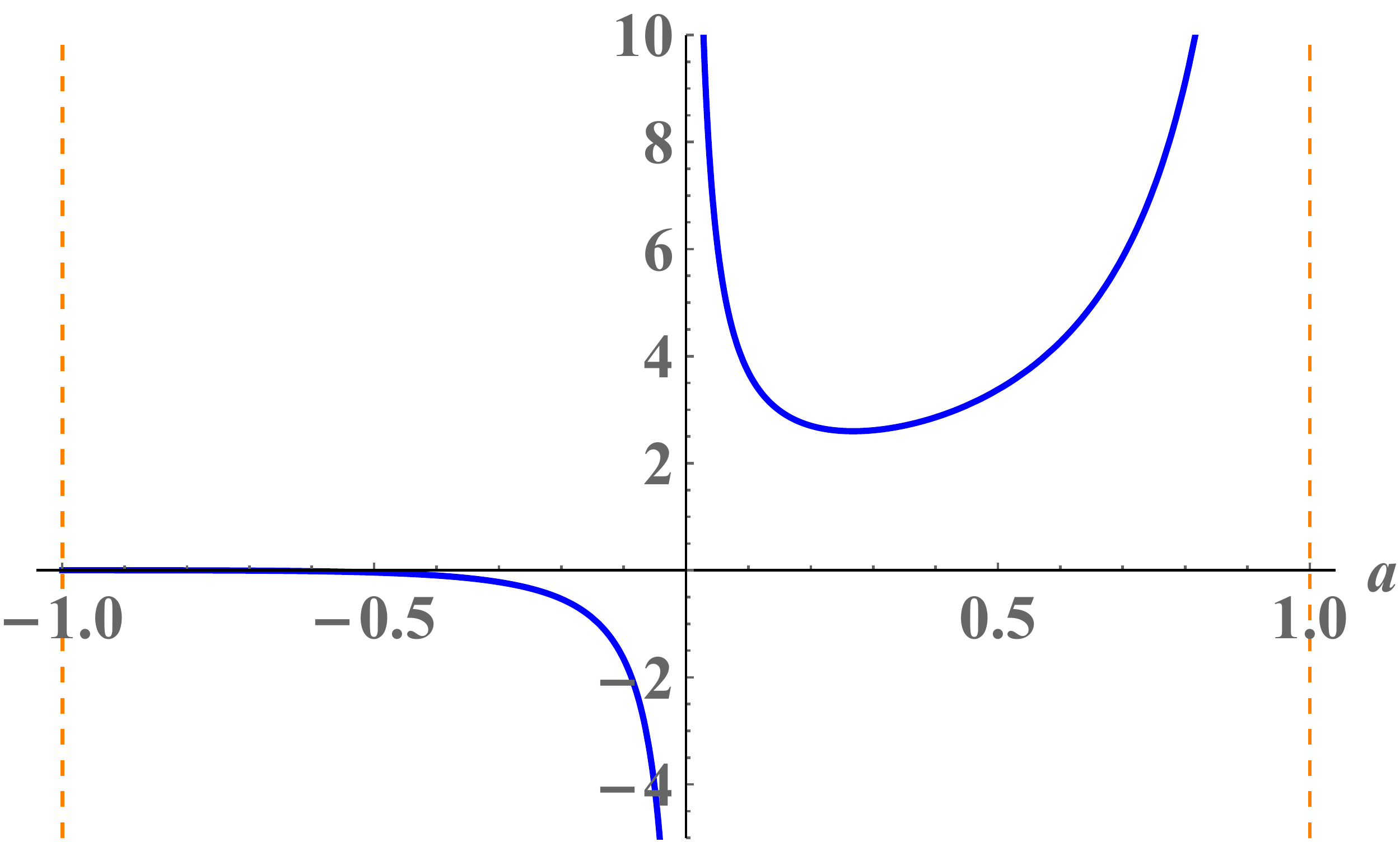}}
\caption{Right-hand side of Eq.\ \eqref{d1kappa} at $-1<a<1$. The minimal positive value $3^{3/2}/2$ is reached at $a_g = 2-\sqrt 3$.}
\label{fig:rhs}
\end{figure}

Below the gap (at $E<E_g$), the DOS \eqref{DOSpsi} is equal to zero, which leads to the condition that $\psi$ is real. The form of solution \eqref{psisol} then implies that $a$ is real and $|a|<1$.
The behavior of the r.h.s.\ of Eq.\ \eqref{d1kappa} in this range of $a$ is shown in Fig.~\ref{fig:rhs}.

The case we are interested in (suppression of $\Delta$ near the surface) corresponds to $d_1>0$ (while $\varkappa$ is real and positive near the gap).
Equation \eqref{d1kappa} then yields two solutions for $a$ at large enough $\varkappa$. They merge and disappear as $\varkappa$ decreases to $\varkappa_g = (3^{1/2}/2^{1/3})d_1^{1/3}$, which determines the dimensionless gap $\varepsilon_g$. The gap value is then given by
\begin{equation} \label{Eg}
\frac{E_g}{\Delta_0} = 1-\frac{3^2}{2^{7/3}} d_1^{4/3},
\end{equation}
which means that the gap in the surface DOS is suppressed in comparison with the bulk value of the order parameter.
Assumption \eqref{kappasmall} implies that $d_1\ll 1$.

We can interpret the result as follows: The spatial scale for the Green function is $\xi_E = \xi_0 /|\varkappa|$, so from the point of view of the spectral gap, information about suppression of $\Delta(x)$ is gathered on this scale. At the same time, $\Delta(x)$ itself is suppressed on much smaller
scale of $r_c$.
Therefore, the effect of $\Delta$ suppression on the gap value will be weakened accordingly:
\begin{equation} \label{Delta0-Eg}
\Delta_0 - E_g \sim \frac 1{\xi_E} \int_0^\infty | \Delta_1 (x) | dx.
\end{equation}
In the dimensionless units, this is written as
\begin{equation} \label{Egestim}
1 - \varepsilon_g \sim \varkappa_g d_1.
\end{equation}
Taking into account Eq.\ \eqref{kappasmall}, we then find $\varkappa_g^3 \sim d_1$, in agreement with Eq.\ (\ref{d1kappa}) and hence with Eq.\ (\ref{Eg}).

Note that the r.h.s.\ of Eq.\ \eqref{Delta0-Eg} can be estimated as $|\Delta_1(0)| r_c/\xi_E$, which is much smaller than $|\Delta_1(0)|$. This implies that the surface suppression of the gap edge [the l.h.s.\ of Eq.\ \eqref{Delta0-Eg}] is much smaller then the surface suppression of the order parameter.

In terms of the discussion presented in Sec.~\ref{sec:intro}, we deal with a shallow Andreev potential well formed near the surface. Impurities smear the Andreev levels out in such a way that the resulting spectral edge $E_g$ is close to the top of the well.

\subsection{Density of states near \texorpdfstring{$E_g$}{Eg}}

According to Eq.\ \eqref{Eg}, deviation of $E_g$ from $\Delta_0$ in dimensionless units is given by
\begin{equation} \label{gamma}
\gamma \equiv 1-\varepsilon_g = 3^2 d_1^{4/3} / 2^{7/3}.
\end{equation}
We want to calculate how finite DOS appears immediately above $E_g$. For that we define dimensionless deviation of $E$ from $E_g$,
\begin{equation}
\epsilon \equiv \varepsilon-\varepsilon_g,
\end{equation}
and consider $\epsilon\ll \gamma$. In this case
\begin{equation}
\varkappa \approx \varkappa_g = ( 2\gamma)^{1/4}.
\end{equation}

Above the gap (at $\epsilon>0$, i.e., at $\varkappa<\varkappa_g$), there are no real solutions of Eq.\ (\ref{d1kappa}) for $a$ (on the physical branch depicted in Fig.~\ref{fig:rhs}), and at $\epsilon \ll \gamma$ we find the following complex solution (the sign is chosen so that the DOS is positive):
\begin{equation}
a = a_g \left( 1 + i \sqrt{3\epsilon/2\gamma} \right).
\end{equation}
This leads to
\begin{equation} \label{psi(ve,X)}
\psi(\varepsilon,X) \approx \psi_0(\varepsilon_g) + 4 \arctanh(a_g e^{-\varkappa_g X}) + \frac{4 (a-a_g) e^{-\varkappa_g X}}{1-a_g^2 e^{-2\varkappa_g X}}
\end{equation}
and
\begin{multline}
\frac{\nu(\epsilon,X)}{\nu_0} = \Im\sinh\psi \approx \frac 12 \Im e^{\psi}
\\
= 2 \sqrt{3} a_g \frac{e^{-\varkappa_g X}(1+a_g e^{-\varkappa_g X})}{(1-a_g e^{-\varkappa_g X})^3} \frac{\sqrt\epsilon}\gamma,
\end{multline}
where we have employed the fact that $\Re\psi \gg 1$ due to $\varepsilon \approx 1$ [$\psi_0(\varepsilon_g)$ in Eq.\ \eqref{psi(ve,X)} is large in this case].
At $x=0$, the expression for the DOS simplifies to
\begin{equation} \label{nunearEg}
\frac{\nu(\epsilon,0)}{\nu_0} = 3\frac{\sqrt\epsilon}\gamma.
\end{equation}

The square-root dependence of the DOS near the spectral edge is characteristic for the mean-field problem of a superconductor with weak magnetic impurities, considered by Abrikosov and Gor'kov (AG) \cite{[][{ [Sov. Phys. JETP  \textbf{12}, 1243 (1961)].}]Abrikosov1960RusEng}, and various other problems that can be mapped
onto it.
In terms of the AG pair-breaking parameter $\eta = 1/\tau_s \Delta_0$ (where $\tau_s$ is the spin-flip scattering time), in the limit of $\eta\ll 1$, the AG result \cite{[][{ [Sov. Phys. JETP  \textbf{12}, 1243 (1961)].}]Abrikosov1960RusEng} for the energy gap corresponds to $\gamma_\mathrm{AG} = 3 \eta^{2/3}/2$, while the relation between $\nu(\epsilon)$ and $\gamma$ has the form
\begin{equation} \label{nunearEgAG}
\frac{\nu_\mathrm{AG}(\epsilon)}{\nu_0} = \sqrt{\frac 32}\frac{\sqrt\epsilon}{\gamma_\mathrm{AG}}.
\end{equation}
Interestingly, our Eq.\ \eqref{nunearEg} differs from this relation by a factor of $\sqrt 6$.

\subsection{Density of states near \texorpdfstring{$\Delta_0$}{Delta0}}

\subsubsection{\texorpdfstring{$E=\Delta_0$}{E=Delta0}}

At $E\to \Delta_0$, the parameter $\varkappa^3/d_1$ in Eq.\ (\ref{d1kappa}) tends to zero, so $a\to -1$. To calculate the DOS at $E=\Delta_0$, we have to keep the correction to this solution. This can be done perturbatively:
\begin{equation} \label{aalpha}
a=-1+\alpha,\qquad |\alpha|\ll 1,
\end{equation}
which immediately yields $\alpha = (-1)^{1/3} \times 2\varkappa/d_1^{1/3}$. There are three possible values of $(-1)^{1/3}$. The real one, $-1$, leads to zero DOS. The complex one producing the positive DOS is
\begin{equation} \label{alpha}
\alpha = 2 e^{i\pi/3} \varkappa / d_1^{1/3}.
\end{equation}

With the help of the identity
\begin{equation}
\arctanh z = \frac 12 \ln\frac{1+z}{1-z},
\end{equation}
the solution (\ref{psisol}) at $x=0$ can be written as
\begin{equation} \label{psi0}
\psi(\varepsilon,0) = \ln \biggl[ \left( \frac{1+a}{1-a} \right)^2 \left( \frac{1+\varepsilon}{1-\varepsilon} \right)^{1/2} \biggr].
\end{equation}
Taking into account Eqs.\ (\ref{aalpha}) and (\ref{alpha}), we obtain
\begin{equation}
\psi(\varepsilon,0)
= \ln ( 2e^{2\pi i/3} / d_1^{2/3} ).
\end{equation}
Since $\Re \psi(\varepsilon,0) \gg 1$, we may write\footnote{Equation \eqref{nuDelta} parametrically coincides with the corresponding AG result
\begin{equation*}
\frac{\nu_\mathrm{AG}(E=\Delta_0)}{\nu_0}
= \frac{3}{2^{11/6} \gamma_\mathrm{AG}^{1/2}},
\end{equation*}
differing only by a numerical factor of $3^{1/2}/2^{1/3}$.}
\begin{equation} \label{nuDelta}
\frac{\nu(E=\Delta_0,0)}{\nu_0}
\approx \frac 12 \Im e^\psi
= \frac{\sqrt 3}{2 d_1^{2/3}}
= \frac{3^{3/2}}{2^{13/6} \gamma^{1/2}}.
\end{equation}

\subsubsection{\texorpdfstring{$E \to \Delta_0$}{E->Delta0}}
\label{sec:EnearDelta0}

Next, we want to find $\nu(E)$ when $E$ deviates slightly from $\Delta_0$. For that, the main-order result \eqref{alpha} in the solution \eqref{aalpha} is not sufficient, and we have to calculate $\alpha$ to higher orders with respect to $\varkappa$ (that encodes the deviation of $E$ from $\Delta_0$; note that $\varkappa$ is real at $E<\Delta_0$ and complex at $E>\Delta_0$). Introducing for brevity
\begin{equation} \label{beta}
\tilde\varkappa \equiv \varkappa/d_1^{1/3},
\end{equation}
we rewrite Eq.\ \eqref{d1kappa} as
\begin{equation} \label{alphaeq}
\alpha^3 = -8\tilde\varkappa^3 (1-\alpha) (1-\alpha/2).
\end{equation}
Its solution at small $\tilde\varkappa$ is expanded into integer powers of $\tilde\varkappa$, and for our calculation the following precision of the perturbation theory is required:
\begin{equation} \label{alphaO}
\alpha = O(\tilde\varkappa) + O(\tilde\varkappa^2) + O(\tilde\varkappa^3).
\end{equation}

Three steps of the perturbation theory for Eq.\ (\ref{alphaeq}) yield\footnote{
In Sec.~\ref{sec:EnearDelta0}, perturbation theory with respect to small $\varkappa$ [or, more precisely, small $\tilde\varkappa$; see Eqs.\ \eqref{beta}--\eqref{alphaO}] is based on Eq.\ \eqref{d1kappa}. Since Eq.\ \eqref{d1kappa} itself is obtained from Eq.\ \eqref{d1kappaunsimplified} at $\varkappa\ll 1$, it is necessary to make sure that no essential contribution is lost within this approach.
One can check that this is indeed so. The reason is that the lost contributions behave as powers of $\varkappa$, while the result that we find contains powers of $\tilde\varkappa$, which is much larger since $d_1\ll 1$.
}
\begin{equation} \label{alphafromM}
\alpha
\approx 2e^{i\pi/3} \tilde\varkappa - 2 e^{2i\pi/3}\tilde\varkappa^2 - 4 \tilde\varkappa^3/3 .
\end{equation}
Plugging this into Eqs.\ \eqref{aalpha}, \eqref{psi0}, and \eqref{DOSpsi}, we find the surface DOS:\footnote{
In the vicinity of $\varepsilon=1$, quantities $(1-\varepsilon)^{1/2}$ and $\varkappa$ are real (positive) at $\varepsilon<1$ and complex at $\varepsilon>1$. In the latter case, the branches of the complex functions should be correctly chosen. It can be checked that in the case of the retarded Green functions that we work with, the correct choice is
\[
(1-\varepsilon)^{1/2} = e^{-i\pi/2} (\varepsilon-1)^{1/2},
\quad
\varkappa = e^{-i\pi/4} (\varepsilon^2-1)^{1/4}.
\]
This refers to Eqs.\ \eqref{UsShG}--\eqref{UsShG0}, \eqref{psisol}, \eqref{d1kappaunsimplified}, \eqref{kappasmall}, \eqref{d1kappa}, \eqref{psi0}, and \eqref{beta}--\eqref{nunearDelta}.
}
\begin{multline} \label{nunearDelta}
\frac{\nu(\varepsilon,0)}{\nu_0} = \frac{\sqrt 3}{2 d_1^{2/3}}
+ \frac{\sqrt 2}{3 d_1^{4/3}} \Im\left[ (1+i\sqrt 3) (1-\varepsilon)^{1/2} \right]
\\
= \frac{3^{3/2}}{2^{13/6} \gamma^{1/2}}
+ \frac{3}{2^{11/6} \gamma} \sqrt{|1-\varepsilon|} \times
\left\{
\begin{array}{cl}
\sqrt 3, & \varepsilon<1,
\\
-1, & \varepsilon>1.
\end{array}
\right.
\end{multline}

The (dimensionless) shift of the spectral edge in the surface DOS corresponds to $(1-\varepsilon)= \gamma$ [see Eq.\ \eqref{gamma}], which sets the natural energy scale for our result \eqref{nunearDelta}. At $(1-\varepsilon) \sim \gamma$, both terms in Eq.\ \eqref{nunearDelta} are of the same order and $\nu(\varepsilon,0) / \nu_0 \sim 1/\gamma^{1/2}$. This can be viewed as moving from $\Delta_0$ towards $E_g$. On the other hand, moving from $E_g$ towards $\Delta_0$, we can apply Eq.\ (\ref{nunearEg}), which yields the same estimate for the DOS at $\epsilon\sim \gamma$. So, the results are consistent and match each other.

\section{Discussion}
\label{sec:numerical}

\begin{figure}
 \center{\includegraphics[width=\columnwidth]{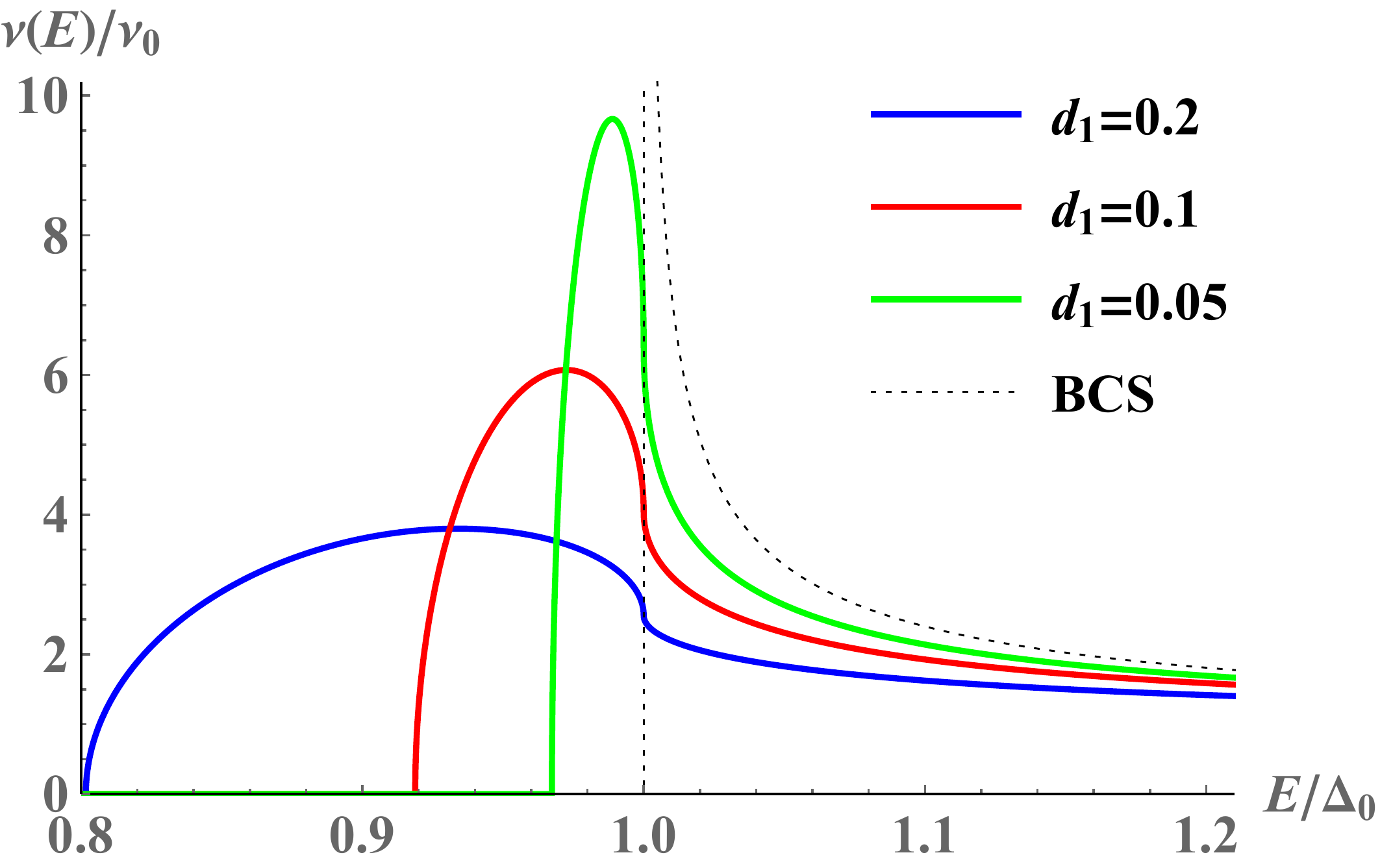}}
\caption{
Surface DOS $\nu(E,0)$ obtained from Eqs.\ \eqref{DOSpsi} and \eqref{psisol} after finding $a$ from Eq.\ \eqref{d1kappaunsimplified} numerically. The colored curves correspond to $d_1=0.2$ (blue), $d_1=0.1$ (red), and $d_1=0.05$ (green). The black dotted curve is the BCS DOS (corresponding to $d_1=0$).
}
\label{fig:DOS}
\end{figure}

We illustrate our results in Fig.~\ref{fig:DOS}, which is obtained by solving Eq.\ \eqref{d1kappaunsimplified} numerically. Although Eq.\ \eqref{d1kappaunsimplified} itself can be considered at arbitrary valued of $d_1$, it was derived and describes our physical system only at $d_1\ll 1$. Therefore, in Fig.~\ref{fig:DOS}, we show the DOS only at small values of $d_1$.

Equation \eqref{d1kappaunsimplified} and hence the curves in Fig.~\ref{fig:DOS} contain information about microscopic parameters of our model only through $d_1$.
As examples of the $\lambda_1(x)$ dependence, we may consider
\begin{equation}
\lambda_1(x) = -|\lambda_1(0)|
\times
\left\{
\begin{array}{ll}
\exp(-x/r_c), & \text{case (a)},
\\
\exp(-x^2/r_c^2), & \text{case (b)}.
\end{array}
\right.
\end{equation}
Assuming $T=0$ for simplicity, from Eq.\ \eqref{d_1} we then find the corresponding results for $d_1$,
\begin{equation}
d_1 = \frac{|\lambda_1(0)|}{\lambda_0^2} \frac{r_c}{\xi_0}
\times
\left\{
\begin{array}{ll}
1, & \text{case (a)},
\\
\sqrt{\pi}/2, & \text{case (b)}.
\end{array}
\right.
\end{equation}

Figure~\ref{fig:DOS} demonstrates suppression of the gap $E_g$ in the surface DOS in comparison with the bulk gap $\Delta_0$; the suppression grows with increasing $d_1$.
Above the gap, the DOS grows as $\sqrt{E-E_g}$,
reaches a maximum at $E_g<E<\Delta_0$, and then decreases passing through the vertical peculiarity at $E=\Delta_0$. At $E>\Delta_0$, the DOS rapidly approaches the BCS result.

The vertical peculiarity is asymmetric. Indeed, according to Eq.\ \eqref{nunearDelta}, the square-root deviation of the DOS from its value at $E=\Delta_0$ has a prefactor that takes different values on the two sides of the peculiarity (on the left, it is $\sqrt 3$ times larger than on the right).

\section{Conclusions}
\label{sec:conclusions}

We have calculated the surface DOS in a superconductor with relatively weak surface suppression of the BCS pairing constant $\lambda(x)$. We are mainly interested in the case of short-range $\lambda(x)$ variation, when its characteristic spatial scale $r_c$ is much smaller than the superconducting coherence length. This case can be experimentally relevant if surface imperfections are limited to the immediate vicinity of the surface. Our main results are analytic and refer to several regions of the $\nu(E)$ dependence.

The gap $E_g$ in the surface DOS differs from the surface value of the order parameter, $\Delta(0)$. With respect to the bulk value of the order parameter, $\Delta_0$, the gap $E_g$ is suppressed much weaker than $\Delta(0)$ [see Eqs.\ \eqref{Eg}--\eqref{Egestim}]. Suppression of $E_g$ with respect to $\Delta_0$ smears the BCS singularity and hence is somewhat similar to the pair breaking considered by Abrikosov and Gor'kov (AG) \cite{[][{ [Sov. Phys. JETP  \textbf{12}, 1243 (1961)].}]Abrikosov1960RusEng}. Similarly to the AG case, $\nu(E) \propto \sqrt{E-E_g}$ immediately above the gap. At the same time, the exact prefactor, being expressed in terms of the gap-edge shift, differs from the AG result by a numerical factor [see Eqs.\ \eqref{nunearEg} and \eqref{nunearEgAG}].

At $E=\Delta_0$, we find a ``vertical'' peculiarity of the DOS,  which implies an infinite-derivative inflection point of the DOS curve.
The value of $\nu$ at $E=\Delta_0$ is large [see Eq.\ \eqref{nuDelta}] and $\nu(E)$ deviates from this value as $\sqrt{|E-\Delta_0|}$ when $E$ deviates from $\Delta_0$. The prefactor of this dependence depends on the sign of $E-\Delta_0$, so the peculiarity is asymmetric [see Eq.\ \eqref{nunearDelta}].

At higher energies, $E>\Delta_0$, the correction to the DOS is found perturbatively.

Experimentally, the surface DOS can be directly probed by scanning tunneling spectroscopy and also directly influences the surface impedance \cite{TinkhamBook,Gurevich2017,Kubo2019}. The zero-temperature threshold for the radiation absorption is given by $2E_g$. This energy determines the threshold behavior of the dissipative conductivity and the surface resistance.

\acknowledgments
The idea of this study was formulated in the course of discussions with M.~V.\ Feigel'man, C.\ Chapelier, and C.\ Tonnoir. We also thank M.~V.\ Feigel'man, M.~A.\ Skvortsov, and K.~S.\ Tikhonov for useful discussions of the results.
Ya.V.F.\ was supported by
the State assignment of the Ministry of Science and Higher Education
and by
the Program of the Russian Academy of Sciences.
A.A.M.\ was partially funded by
the RFBR research projects 18-02-00318
and 18-52-45011-IND.


\appendix

\section{Fluctuation propagator}
\label{sec:app:L0}

The static propagator of superconducting fluctuations, $L_0(k)$, is defined by the following relations \cite{[][{ [JETP  \textbf{117}, 487 (2013)].}]Skvortsov2013RusEng}:
\begin{multline} \label{L0inv}
L_0^{-1}(k) = \pi T \sum_n \biggl( \frac{\sin\theta_0}{\Delta_0} - \frac{\cos^2 \theta_0}{\frac D2 k^2 + \omega_n \cos\theta_0 + \Delta_0 \sin\theta_0} \biggr)
\\
= 2\pi \frac T{\Delta_0} \sum_{n= 0}^\infty \frac{1 +k^2 \xi_0^2 \sqrt{(\omega_n/\Delta_0)^2+1}}{\left[ (\omega_n /\Delta_0)^2 +1 \right] (k^2 \xi_0^2 +\sqrt{(\omega_n/\Delta_0)^2+1})}.
\end{multline}
The sum over the Matsubara frequencies here cannot be calculated in the general case, and we now consider some important limiting cases.

At zero temperature ($T=0$), the Matsubara sum in Eq.\ (\ref{L0inv}) is substituted by the integral, which can be calculated and written in terms of $K \equiv k \xi_0$ as \cite{Meyer2001,[][{ [JETP  \textbf{117}, 487 (2013)].}]Skvortsov2013RusEng}
\begin{equation}
L_0^{-1}(k) = \frac{\pi}{2K^2} + \frac{\sqrt{K^4-1}}{K^2} \ln\left( K^2+ \sqrt{K^4-1} \right).
\end{equation}

Near the critical temperature ($T\to T_c$), we may put $[(\omega_n/\Delta_0)^2+1]\approx (\omega_n/\Delta_0)^2$ in Eq.\ (\ref{L0inv}), and then the sum can be calculated:
\begin{multline}
L_0^{-1}(k)
= \frac{\pi \Delta_0(T)}{4 K^2 T_c}
\\
+ \left( 1-\frac 1{K^4} \right)
\left[ \psi\left( \frac 12 +\frac{K^2 \Delta_0(T)}{2\pi T_c} \right) - \psi\left( \frac 12 \right) \right],
\end{multline}
where $\psi$ is the digamma function. The temperature dependence of the order parameter near $T_c$ is given by \cite{AbrikosovBookEng}
\begin{equation} \label{Delta0nearTc}
\Delta_0(T) = \pi \sqrt{\frac{8}{7\zeta(3)}} \sqrt{T_c(T_c-T)}.
\end{equation}

At $k=0$, considering $L_0^{-1}(0)$ as a function of temperature, we find
\begin{equation}
L_0^{-1}(0) =
\left\{
\begin{array}{ll}
1, & T\ll T_c,
\\
\frac{7\zeta(3) \Delta_0^2(T)}{4\pi^2 T_c^2}, & (T_c-T) \ll T_c.
\end{array}
\right.
\end{equation}
With the help of Eq.\ \eqref{Delta0nearTc}, the result at $(T_c-T)\ll T_c$ can be written as
\begin{equation}
L_0^{-1}(0)
=2 (1- T/T_c).
\end{equation}

\section{Applicability of nonperturbative results}
\label{sec:app:applnonpert}

The results of Sec.~\ref{sec:DOSnonperturb} require conditions \eqref{rcllxi0} and \eqref{kappasmall} to be satisfied (meaning that the energies $E$ considered are close enough to $\Delta_0$).
The conditions can be summarized as
\begin{equation}
r_c \ll \xi_0 \ll \xi_E.
\end{equation}

The first condition, $r_c \ll \xi_0$, is formulated in terms of the input parameters of our model (small spatial scale of the pairing-constant variations).
However, the second condition $\xi_0 \ll \xi_E$, depends on the energy $E$ that we consider.
It becomes most restrictive at $E=E_g$. Our result \eqref{Eg} thus implies that it is sufficient to require condition  \eqref{d1ll1}.

Since the spatial scale for $\Delta_1(x)$ is $r_c$, with the help of the definition of $d_1$ in Eq.\ (\ref{d1}), we can rewrite condition \eqref{d1ll1} as
\begin{equation} \label{applcond1}
\frac{| \Delta_1(0) |}{\Delta_0} \frac{r_c}{\xi_0} \ll 1.
\end{equation}

In terms of $\lambda_1(x)$, the $d_1$ parameter is given by Eq.\ (\ref{d_1}).
At $T=0$, we have $L_0(0) = 1$, and Eq.\ (\ref{d_1}) allows us to rewrite condition \eqref{d1ll1} in terms of the input parameters of our model as
\begin{equation} \label{applcond2}
\frac{| \lambda_1(0) |}{\lambda_0^2} \frac{r_c}{\xi_0} \ll 1.
\end{equation}

At the same time, during construction of the self-consistent perturbation theory in Sec.~\ref{sec:self-cons_perturb}, conditions
\begin{equation} \label{applcond}
  | \Delta_1(0) | / \Delta_0 \ll 1,
  \qquad
  | \lambda_1(0) | / \lambda_0^2 \ll 1
\end{equation}
had to be satisfied. Then Eqs.\ \eqref{applcond1} and \eqref{applcond2} do not add anything new.

The applicability conditions for the results of Sec.~\ref{sec:DOSnonperturb} are therefore given by Eqs.\ \eqref{rcllxi0} and \eqref{applcond}, while condition \eqref{d1ll1} is their direct consequence.


\begin{thebibliography}{27}%
\makeatletter
\providecommand \@ifxundefined [1]{%
 \@ifx{#1\undefined}
}%
\providecommand \@ifnum [1]{%
 \ifnum #1\expandafter \@firstoftwo
 \else \expandafter \@secondoftwo
 \fi
}%
\providecommand \@ifx [1]{%
 \ifx #1\expandafter \@firstoftwo
 \else \expandafter \@secondoftwo
 \fi
}%
\providecommand \natexlab [1]{#1}%
\providecommand \enquote  [1]{``#1''}%
\providecommand \bibnamefont  [1]{#1}%
\providecommand \bibfnamefont [1]{#1}%
\providecommand \citenamefont [1]{#1}%
\providecommand \href@noop [0]{\@secondoftwo}%
\providecommand \href [0]{\begingroup \@sanitize@url \@href}%
\providecommand \@href[1]{\@@startlink{#1}\@@href}%
\providecommand \@@href[1]{\endgroup#1\@@endlink}%
\providecommand \@sanitize@url [0]{\catcode `\\12\catcode `\$12\catcode
  `\&12\catcode `\#12\catcode `\^12\catcode `\_12\catcode `\%12\relax}%
\providecommand \@@startlink[1]{}%
\providecommand \@@endlink[0]{}%
\providecommand \url  [0]{\begingroup\@sanitize@url \@url }%
\providecommand \@url [1]{\endgroup\@href {#1}{\urlprefix }}%
\providecommand \urlprefix  [0]{URL }%
\providecommand \Eprint [0]{\href }%
\providecommand \doibase [0]{https://doi.org/}%
\providecommand \selectlanguage [0]{\@gobble}%
\providecommand \bibinfo  [0]{\@secondoftwo}%
\providecommand \bibfield  [0]{\@secondoftwo}%
\providecommand \translation [1]{[#1]}%
\providecommand \BibitemOpen [0]{}%
\providecommand \bibitemStop [0]{}%
\providecommand \bibitemNoStop [0]{.\EOS\space}%
\providecommand \EOS [0]{\spacefactor3000\relax}%
\providecommand \BibitemShut  [1]{\csname bibitem#1\endcsname}%
\let\auto@bib@innerbib\@empty
\bibitem [{\citenamefont {Bardeen}\ \emph {et~al.}(1957)\citenamefont
  {Bardeen}, \citenamefont {Cooper},\ and\ \citenamefont
  {Schrieffer}}]{Bardeen1957b}%
  \BibitemOpen
  \bibfield  {author} {\bibinfo {author} {\bibfnamefont {J.}~\bibnamefont
  {Bardeen}}, \bibinfo {author} {\bibfnamefont {L.~N.}\ \bibnamefont
  {Cooper}},\ and\ \bibinfo {author} {\bibfnamefont {J.~R.}\ \bibnamefont
  {Schrieffer}},\ }\bibfield  {title} {\bibinfo {title} {Theory of
  superconductivity},\ }\href@noop {} {\bibfield  {journal} {\bibinfo
  {journal} {Phys. Rev.}\ }\textbf {\bibinfo {volume} {108}},\ \bibinfo {pages}
  {1175} (\bibinfo {year} {1957})}\BibitemShut {NoStop}%
\bibitem [{\citenamefont {Tinkham}(2004)}]{TinkhamBook}%
  \BibitemOpen
  \bibfield  {author} {\bibinfo {author} {\bibfnamefont {M.}~\bibnamefont
  {Tinkham}},\ }\href@noop {} {\emph {\bibinfo {title} {Introduction to
  Superconductivity (2nd edition)}}}\ (\bibinfo  {publisher} {Dover},\ \bibinfo
  {address} {New York},\ \bibinfo {year} {2004})\BibitemShut {NoStop}%
\bibitem [{\citenamefont {Andreev}(1964)}]{Andreev1964RusEng}%
  \BibitemOpen
  \bibfield  {author} {\bibinfo {author} {\bibfnamefont {A.~F.}\ \bibnamefont
  {Andreev}},\ }\bibfield  {title} {\bibinfo {title} {Thermal conductivity of
  the intermediate state of superconductors},\ }\href@noop {} {\bibfield
  {journal} {\bibinfo  {journal} {Zh. Eksp. Teor. Fiz.}\ }\textbf {\bibinfo
  {volume} {46}},\ \bibinfo {pages} {1823} (\bibinfo {year}
  {1964})}\BibitemShut {NoStop}%
\bibitem [{\citenamefont {Shnirman}\ \emph {et~al.}(1999)\citenamefont
  {Shnirman}, \citenamefont {Adagideli}, \citenamefont {Goldbart},\ and\
  \citenamefont {Yazdani}}]{Shnirman1999.PhysRevB.60.7517}%
  \BibitemOpen
  \bibfield  {author} {\bibinfo {author} {\bibfnamefont {A.}~\bibnamefont
  {Shnirman}}, \bibinfo {author} {\bibfnamefont {{\.I}.}~\bibnamefont
  {Adagideli}}, \bibinfo {author} {\bibfnamefont {P.~M.}\ \bibnamefont
  {Goldbart}},\ and\ \bibinfo {author} {\bibfnamefont {A.}~\bibnamefont
  {Yazdani}},\ }\bibfield  {title} {\bibinfo {title} {Resonant states and
  order-parameter suppression near pointlike impurities in d-wave
  superconductors},\ }\href {https://doi.org/10.1103/PhysRevB.60.7517}
  {\bibfield  {journal} {\bibinfo  {journal} {Phys. Rev. B}\ }\textbf {\bibinfo
  {volume} {60}},\ \bibinfo {pages} {7517} (\bibinfo {year}
  {1999})}\BibitemShut {NoStop}%
\bibitem [{\citenamefont {Andersen}\ \emph {et~al.}(2006)\citenamefont
  {Andersen}, \citenamefont {Melikyan}, \citenamefont {Nunner},\ and\
  \citenamefont {Hirschfeld}}]{Andersen2006.PhysRevLett.96.097004}%
  \BibitemOpen
  \bibfield  {author} {\bibinfo {author} {\bibfnamefont {B.~M.}\ \bibnamefont
  {Andersen}}, \bibinfo {author} {\bibfnamefont {A.}~\bibnamefont {Melikyan}},
  \bibinfo {author} {\bibfnamefont {T.~S.}\ \bibnamefont {Nunner}},\ and\
  \bibinfo {author} {\bibfnamefont {P.~J.}\ \bibnamefont {Hirschfeld}},\
  }\bibfield  {title} {\bibinfo {title} {Andreev states near short-ranged
  pairing potential impurities},\ }\href
  {https://doi.org/10.1103/PhysRevLett.96.097004} {\bibfield  {journal}
  {\bibinfo  {journal} {Phys. Rev. Lett.}\ }\textbf {\bibinfo {volume} {96}},\
  \bibinfo {pages} {097004} (\bibinfo {year} {2006})}\BibitemShut {NoStop}%
\bibitem [{\citenamefont {Bespalov}(2019)}]{Bespalov2019.PhysRevB.100.094507}%
  \BibitemOpen
  \bibfield  {author} {\bibinfo {author} {\bibfnamefont {A.~A.}\ \bibnamefont
  {Bespalov}},\ }\bibfield  {title} {\bibinfo {title} {Impurity-induced subgap
  states in superconductors with inhomogeneous pairing},\ }\href
  {https://doi.org/10.1103/PhysRevB.100.094507} {\bibfield  {journal} {\bibinfo
   {journal} {Phys. Rev. B}\ }\textbf {\bibinfo {volume} {100}},\ \bibinfo
  {pages} {094507} (\bibinfo {year} {2019})}\BibitemShut {NoStop}%
\bibitem [{\citenamefont {Martin}\ \emph {et~al.}(2005)\citenamefont {Martin},
  \citenamefont {Podolsky},\ and\ \citenamefont
  {Kivelson}}]{Martin2005.PhysRevB.72.060502}%
  \BibitemOpen
  \bibfield  {author} {\bibinfo {author} {\bibfnamefont {I.}~\bibnamefont
  {Martin}}, \bibinfo {author} {\bibfnamefont {D.}~\bibnamefont {Podolsky}},\
  and\ \bibinfo {author} {\bibfnamefont {S.~A.}\ \bibnamefont {Kivelson}},\
  }\bibfield  {title} {\bibinfo {title} {Enhancement of superconductivity by
  local inhomogeneities},\ }\href {https://doi.org/10.1103/PhysRevB.72.060502}
  {\bibfield  {journal} {\bibinfo  {journal} {Phys. Rev. B}\ }\textbf {\bibinfo
  {volume} {72}},\ \bibinfo {pages} {060502} (\bibinfo {year}
  {2005})}\BibitemShut {NoStop}%
\bibitem [{\citenamefont {Zou}\ \emph {et~al.}(2008)\citenamefont {Zou},
  \citenamefont {Klich},\ and\ \citenamefont
  {Refael}}]{Zou2008.PhysRevB.77.144523}%
  \BibitemOpen
  \bibfield  {author} {\bibinfo {author} {\bibfnamefont {Y.}~\bibnamefont
  {Zou}}, \bibinfo {author} {\bibfnamefont {I.}~\bibnamefont {Klich}},\ and\
  \bibinfo {author} {\bibfnamefont {G.}~\bibnamefont {Refael}},\ }\bibfield
  {title} {\bibinfo {title} {Effect of inhomogeneous coupling on {BCS}
  superconductors},\ }\href {https://doi.org/10.1103/PhysRevB.77.144523}
  {\bibfield  {journal} {\bibinfo  {journal} {Phys. Rev. B}\ }\textbf {\bibinfo
  {volume} {77}},\ \bibinfo {pages} {144523} (\bibinfo {year}
  {2008})}\BibitemShut {NoStop}%
\bibitem [{\citenamefont {R\o{}mer}\ \emph {et~al.}(2012)\citenamefont
  {R\o{}mer}, \citenamefont {Graser}, \citenamefont {Nunner}, \citenamefont
  {Hirschfeld},\ and\ \citenamefont {Andersen}}]{Romer2012.PhysRevB.86.054507}%
  \BibitemOpen
  \bibfield  {author} {\bibinfo {author} {\bibfnamefont {A.~T.}\ \bibnamefont
  {R\o{}mer}}, \bibinfo {author} {\bibfnamefont {S.}~\bibnamefont {Graser}},
  \bibinfo {author} {\bibfnamefont {T.~S.}\ \bibnamefont {Nunner}}, \bibinfo
  {author} {\bibfnamefont {P.~J.}\ \bibnamefont {Hirschfeld}},\ and\ \bibinfo
  {author} {\bibfnamefont {B.~M.}\ \bibnamefont {Andersen}},\ }\bibfield
  {title} {\bibinfo {title} {Local modulations of the spin-fluctuation-mediated
  pairing interaction by impurities in $d$-wave superconductors},\ }\href
  {https://doi.org/10.1103/PhysRevB.86.054507} {\bibfield  {journal} {\bibinfo
  {journal} {Phys. Rev. B}\ }\textbf {\bibinfo {volume} {86}},\ \bibinfo
  {pages} {054507} (\bibinfo {year} {2012})}\BibitemShut {NoStop}%
\bibitem [{\citenamefont {Golubov}\ and\ \citenamefont
  {Kupriyanov}(1989)}]{Golubov1989RusEng}%
  \BibitemOpen
  \bibfield  {author} {\bibinfo {author} {\bibfnamefont {A.~A.}\ \bibnamefont
  {Golubov}}\ and\ \bibinfo {author} {\bibfnamefont {M.~{\relax Yu}.}\
  \bibnamefont {Kupriyanov}},\ }\bibfield  {title} {\bibinfo {title} {Josephson
  effect in {SNINS} and {SNIS} tunnel structures with finite transparency of
  the {SN} boundaries},\ }\href@noop {} {\bibfield  {journal} {\bibinfo
  {journal} {Zh. Eksp. Teor. Fiz.}\ }\textbf {\bibinfo {volume} {96}},\
  \bibinfo {pages} {1420} (\bibinfo {year} {1989})}\BibitemShut {NoStop}%
\bibitem [{\citenamefont {Zhou}\ \emph {et~al.}(1998)\citenamefont {Zhou},
  \citenamefont {Charlat}, \citenamefont {Spivak},\ and\ \citenamefont
  {Pannetier}}]{Zhou1998}%
  \BibitemOpen
  \bibfield  {author} {\bibinfo {author} {\bibfnamefont {F.}~\bibnamefont
  {Zhou}}, \bibinfo {author} {\bibfnamefont {P.}~\bibnamefont {Charlat}},
  \bibinfo {author} {\bibfnamefont {B.}~\bibnamefont {Spivak}},\ and\ \bibinfo
  {author} {\bibfnamefont {B.}~\bibnamefont {Pannetier}},\ }\bibfield  {title}
  {\bibinfo {title} {Density of states in superconductor--normal
  metal--superconductor junctions},\ }\href@noop {} {\bibfield  {journal}
  {\bibinfo  {journal} {J. Low Temp. Phys.}\ }\textbf {\bibinfo {volume}
  {110}},\ \bibinfo {pages} {841} (\bibinfo {year} {1998})}\BibitemShut
  {NoStop}%
\bibitem [{\citenamefont {Antoine}(2012)}]{AntoineBook}%
  \BibitemOpen
  \bibfield  {author} {\bibinfo {author} {\bibfnamefont {C.~Z.}\ \bibnamefont
  {Antoine}},\ }\href@noop {} {\emph {\bibinfo {title} {Materials and Surface
  Aspects in the Development of SRF Niobium Cavities}}}\ (\bibinfo  {publisher}
  {Institute of Electronic Systems},\ \bibinfo {address} {Warsaw University of
  Technology},\ \bibinfo {year} {2012})\BibitemShut {NoStop}%
\bibitem [{\citenamefont {Gurevich}(2012)}]{Gurevich2012}%
  \BibitemOpen
  \bibfield  {author} {\bibinfo {author} {\bibfnamefont {A.}~\bibnamefont
  {Gurevich}},\ }\bibfield  {title} {\bibinfo {title} {Superconducting
  radio-frequency fundamentals for particle accelerators},\ }\href@noop {}
  {\bibfield  {journal} {\bibinfo  {journal} {Rev. Accel. Sci. Technol.}\
  }\textbf {\bibinfo {volume} {5}},\ \bibinfo {pages} {119} (\bibinfo {year}
  {2012})}\BibitemShut {NoStop}%
\bibitem [{\citenamefont {Halama}(1971)}]{Halama1971}%
  \BibitemOpen
  \bibfield  {author} {\bibinfo {author} {\bibfnamefont {H.~J.}\ \bibnamefont
  {Halama}},\ }\bibfield  {title} {\bibinfo {title} {Effects of radiation on
  surface resistance of superconducting niobium cavity},\ }\href@noop {}
  {\bibfield  {journal} {\bibinfo  {journal} {Appl. Phys. Lett.}\ }\textbf
  {\bibinfo {volume} {19}},\ \bibinfo {pages} {90} (\bibinfo {year}
  {1971})}\BibitemShut {NoStop}%
\bibitem [{\citenamefont {Mazanik}(2016)}]{MazanikBSThesis2016}%
  \BibitemOpen
  \bibfield  {author} {\bibinfo {author} {\bibfnamefont {A.~A.}\ \bibnamefont
  {Mazanik}},\ }\href@noop {} {\emph {\bibinfo {title} {Density of states at
  the surface of a superconductor with a nonuniform coupling constant (in
  Russian)}}}\ (\bibinfo  {publisher} {Bachelor's Thesis},\ \bibinfo {address}
  {MIPT},\ \bibinfo {year} {2016})\ \bibinfo {note}
  {\href{http://chair.itp.ac.ru/biblio/bachelors/2016/mazanik_bak_diplom_2016.pdf}{http://chair.itp.ac.ru}}\BibitemShut
  {NoStop}%
\bibitem [{\citenamefont {Razumovskiy}(2017)}]{RazumovskiyBSThesis2017}%
  \BibitemOpen
  \bibfield  {author} {\bibinfo {author} {\bibfnamefont {M.~V.}\ \bibnamefont
  {Razumovskiy}},\ }\href@noop {} {\emph {\bibinfo {title} {Surface density of
  states near the spectrum edge in superconductors with inhomogeneous coupling
  constant (in Russian)}}}\ (\bibinfo  {publisher} {Bachelor's Thesis},\
  \bibinfo {address} {MIPT},\ \bibinfo {year} {2017})\ \bibinfo {note}
  {\href{http://chair.itp.ac.ru/biblio/bachelors/2017/razumovskii_bak_diplom_2017.pdf}{http://chair.itp.ac.ru}}\BibitemShut
  {NoStop}%
\bibitem [{\citenamefont {Gurevich}\ and\ \citenamefont
  {Kubo}(2017)}]{Gurevich2017}%
  \BibitemOpen
  \bibfield  {author} {\bibinfo {author} {\bibfnamefont {A.}~\bibnamefont
  {Gurevich}}\ and\ \bibinfo {author} {\bibfnamefont {T.}~\bibnamefont
  {Kubo}},\ }\bibfield  {title} {\bibinfo {title} {Surface impedance and
  optimum surface resistance of a superconductor with an imperfect surface},\
  }\href@noop {} {\bibfield  {journal} {\bibinfo  {journal} {Phys. Rev. B}\
  }\textbf {\bibinfo {volume} {96}},\ \bibinfo {pages} {184515} (\bibinfo
  {year} {2017})}\BibitemShut {NoStop}%
\bibitem [{\citenamefont {Kubo}\ and\ \citenamefont
  {Gurevich}(2019)}]{Kubo2019}%
  \BibitemOpen
  \bibfield  {author} {\bibinfo {author} {\bibfnamefont {T.}~\bibnamefont
  {Kubo}}\ and\ \bibinfo {author} {\bibfnamefont {A.}~\bibnamefont
  {Gurevich}},\ }\bibfield  {title} {\bibinfo {title} {Field-dependent
  nonlinear surface resistance and its optimization by surface nanostructuring
  in superconductors},\ }\href@noop {} {\bibfield  {journal} {\bibinfo
  {journal} {Phys. Rev. B}\ }\textbf {\bibinfo {volume} {100}},\ \bibinfo
  {pages} {064522} (\bibinfo {year} {2019})}\BibitemShut {NoStop}%
\bibitem [{\citenamefont {Noffsinger}\ and\ \citenamefont
  {Cohen}(2010)}]{Noffsinger2010.PhysRevB.81.214519}%
  \BibitemOpen
  \bibfield  {author} {\bibinfo {author} {\bibfnamefont {J.}~\bibnamefont
  {Noffsinger}}\ and\ \bibinfo {author} {\bibfnamefont {M.~L.}\ \bibnamefont
  {Cohen}},\ }\bibfield  {title} {\bibinfo {title} {First-principles
  calculation of the electron-phonon coupling in ultrathin {Pb}
  superconductors: {S}uppression of the transition temperature by surface
  phonons},\ }\href {https://doi.org/10.1103/PhysRevB.81.214519} {\bibfield
  {journal} {\bibinfo  {journal} {Phys. Rev. B}\ }\textbf {\bibinfo {volume}
  {81}},\ \bibinfo {pages} {214519} (\bibinfo {year} {2010})}\BibitemShut
  {NoStop}%
\bibitem [{\citenamefont {Ginzburg}(1964)}]{Ginzburg1964}%
  \BibitemOpen
  \bibfield  {author} {\bibinfo {author} {\bibfnamefont {V.~L.}\ \bibnamefont
  {Ginzburg}},\ }\bibfield  {title} {\bibinfo {title} {On surface
  superconductivity},\ }\href {https://doi.org/10.1016/0031-9163(64)90672-9}
  {\bibfield  {journal} {\bibinfo  {journal} {Physics Letters}\ }\textbf
  {\bibinfo {volume} {13}},\ \bibinfo {pages} {101} (\bibinfo {year}
  {1964})}\BibitemShut {NoStop}%
\bibitem [{\citenamefont {Larkin}\ and\ \citenamefont
  {Ovchinnikov}(1971)}]{Larkin1971RusEng}%
  \BibitemOpen
  \bibfield  {author} {\bibinfo {author} {\bibfnamefont {A.~I.}\ \bibnamefont
  {Larkin}}\ and\ \bibinfo {author} {\bibfnamefont {{\relax Yu}.~N.}\
  \bibnamefont {Ovchinnikov}},\ }\bibfield  {title} {\bibinfo {title} {Density
  of states in inhomogeneous superconductors},\ }\href@noop {} {\bibfield
  {journal} {\bibinfo  {journal} {Zh. Eksp. Teor. Fiz.}\ }\textbf {\bibinfo
  {volume} {61}},\ \bibinfo {pages} {2147} (\bibinfo {year}
  {1971})}\BibitemShut {NoStop}%
\bibitem [{\citenamefont {Meyer}\ and\ \citenamefont
  {Simons}(2001)}]{Meyer2001}%
  \BibitemOpen
  \bibfield  {author} {\bibinfo {author} {\bibfnamefont {J.~S.}\ \bibnamefont
  {Meyer}}\ and\ \bibinfo {author} {\bibfnamefont {B.~D.}\ \bibnamefont
  {Simons}},\ }\bibfield  {title} {\bibinfo {title} {Gap fluctuations in
  inhomogeneous superconductors},\ }\href@noop {} {\bibfield  {journal}
  {\bibinfo  {journal} {Phys. Rev. B}\ }\textbf {\bibinfo {volume} {64}},\
  \bibinfo {pages} {134516} (\bibinfo {year} {2001})}\BibitemShut {NoStop}%
\bibitem [{\citenamefont {Skvortsov}\ and\ \citenamefont
  {Feigel'man}(2013)}]{Skvortsov2013RusEng}%
  \BibitemOpen
  \bibfield  {author} {\bibinfo {author} {\bibfnamefont {M.~A.}\ \bibnamefont
  {Skvortsov}}\ and\ \bibinfo {author} {\bibfnamefont {M.~V.}\ \bibnamefont
  {Feigel'man}},\ }\bibfield  {title} {\bibinfo {title} {Subgap states in
  disordered superconductors},\ }\href@noop {} {\bibfield  {journal} {\bibinfo
  {journal} {Zh. Eksp. Teor. Fiz.}\ }\textbf {\bibinfo {volume} {144}},\
  \bibinfo {pages} {560} (\bibinfo {year} {2013})}\BibitemShut {NoStop}%
\bibitem [{\citenamefont {Usadel}(1970)}]{Usadel1970}%
  \BibitemOpen
  \bibfield  {author} {\bibinfo {author} {\bibfnamefont {K.~D.}\ \bibnamefont
  {Usadel}},\ }\bibfield  {title} {\bibinfo {title} {Generalized diffusion
  equation for superconducting alloys},\ }\href@noop {} {\bibfield  {journal}
  {\bibinfo  {journal} {Phys. Rev. Lett.}\ }\textbf {\bibinfo {volume} {25}},\
  \bibinfo {pages} {507} (\bibinfo {year} {1970})}\BibitemShut {NoStop}%
\bibitem [{\citenamefont {Belzig}\ \emph {et~al.}(1999)\citenamefont {Belzig},
  \citenamefont {Wilhelm}, \citenamefont {Bruder}, \citenamefont {Sch{\"o}n},\
  and\ \citenamefont {Zaikin}}]{Belzig1999}%
  \BibitemOpen
  \bibfield  {author} {\bibinfo {author} {\bibfnamefont {W.}~\bibnamefont
  {Belzig}}, \bibinfo {author} {\bibfnamefont {F.~K.}\ \bibnamefont {Wilhelm}},
  \bibinfo {author} {\bibfnamefont {C.}~\bibnamefont {Bruder}}, \bibinfo
  {author} {\bibfnamefont {G.}~\bibnamefont {Sch{\"o}n}},\ and\ \bibinfo
  {author} {\bibfnamefont {A.~D.}\ \bibnamefont {Zaikin}},\ }\bibfield  {title}
  {\bibinfo {title} {Quasiclassical {Green}'s function approach to mesoscopic
  superconductivity},\ }\href@noop {} {\bibfield  {journal} {\bibinfo
  {journal} {Superlattices Microstruct.}\ }\textbf {\bibinfo {volume} {25}},\
  \bibinfo {pages} {1251} (\bibinfo {year} {1999})}\BibitemShut {NoStop}%
\bibitem [{\citenamefont {Abrikosov}\ and\ \citenamefont
  {Gor'kov}(1960)}]{Abrikosov1960RusEng}%
  \BibitemOpen
  \bibfield  {author} {\bibinfo {author} {\bibfnamefont {A.~A.}\ \bibnamefont
  {Abrikosov}}\ and\ \bibinfo {author} {\bibfnamefont {L.~P.}\ \bibnamefont
  {Gor'kov}},\ }\bibfield  {title} {\bibinfo {title} {Contribution to the
  theory of superconducting alloys with paramagnetic impurities},\ }\href@noop
  {} {\bibfield  {journal} {\bibinfo  {journal} {Zh. Eksp. Teor. Fiz.}\
  }\textbf {\bibinfo {volume} {39}},\ \bibinfo {pages} {1781} (\bibinfo {year}
  {1960})}\BibitemShut {NoStop}%
\bibitem [{\citenamefont {Abrikosov}(1988)}]{AbrikosovBookEng}%
  \BibitemOpen
  \bibfield  {author} {\bibinfo {author} {\bibfnamefont {A.~A.}\ \bibnamefont
  {Abrikosov}},\ }\href@noop {} {\emph {\bibinfo {title} {Fundamentals of the
  Theory of Metals}}}\ (\bibinfo  {publisher} {North-Holland},\ \bibinfo
  {address} {Amsterdam},\ \bibinfo {year} {1988})\BibitemShut {NoStop}%
\end{thebibliography}
%

\end{document}